\documentclass[twocolumn,amsmath,floats,showpacs,nofootinbib]{revtex4}
\usepackage[usenames]{color}
\usepackage{graphicx}
\usepackage{epstopdf}
\usepackage{textcomp}
\usepackage{hyperref}
\usepackage{multirow}
\usepackage{tikz}
\usepackage{rotating}

\hypersetup{colorlinks=true}

\begin{document}

\title{Point-contact studies of high-temperature superconductor $\rm YBa_2Cu_3O_{7-\delta}$}

\author{I. K. Yanson, L. F. Rybal'chenko,  V. V. Fisun, N. L. Bobrov, V.M.~Kirzhner, Yu. D. Tret'yakov*, A. R. Kaul'*, and I. E. Graboi*}
\affiliation{Physicotechnical Institute of Low Temperatures, Academy of Sciences of the Ukrainian SSR, Kharkov,\\
A. M. Gorky State University, Kharkov,\\
and M. V. Lomonosov State University, Moscow*\\
Email address: bobrov@ilt.kharkov.ua}
\published {(\href{http://fntr.ilt.kharkov.ua/fnt/pdf/14/14-7/f14-0732r.pdf}{Fiz. Nizk. Temp.} , \textbf{14}, 732 (1988)); (Sov. J. Low Temp. Phys., \textbf{14}, 402 (1988)}
\date{\today}

\begin{abstract}Point-contacts formed by the high-temperature superconductor $\rm YBa_2Cu_3O_{7-\delta}$ prepared by cryochemical technique, and a noble metal Ag or Cu) are investigated. The maximum value of the energy gap $\Delta\simeq 40~meV$ and the ratio $2\Delta/kT_c\simeq 12$ are obtained. It is found that along certain crystallographic axes of the superconductor under investigation, the electrical resistivity $\lesssim 10^{-5} \Omega\cdot cm$.

\pacs {74.45.+c,  73.40.-c,  74.20.Mn,  74.72.-h, 74.72.Bk}

\end{abstract}

\maketitle

The existence of an energy gap in new metal oxide superconductors has been established experimentally beyond any doubt. This conclusion was drawn, for example, from the investigations of junctions with weak coupling, both of tunnel type \cite{1} and with a direct conductivity (point-contact type) \cite{2}. However, there are considerable discrepancies in the results obtained by different groups on the energy gap and its temperature dependence.

In this paper, we present the results of point-contact (PC) studies in the metal oxide superconductor $\rm YBa_2Cu_3O_{7-\delta}$ prepared by using the cryochemical technique \cite{3} involving the dispersion of an aqueous solution of ceramic forming components into liquid nitrogen, dehydration of an ice-and-salt mixture by sublimation, thermal decomposition of salts, and fritting of a ceramic followed by its cooling in an atmosphere of oxygen.

The x-ray phase analysis on the set-up DRON-2 ($\rm Cu$ $K_{\alpha}$-emission in $\rm Cu$) was carried out on singlephase samples characterized by a rhombic structure and lattice parameters $a=3.836\pm 0.006; b = 3.888 \pm 0.002; c = 11.70 \pm 0.01$. Chemical analysis based on the determination of the oxidation level of copper by iodometric titration of a sample dissolved in $\rm HCl$ containing $\rm KI$ showed that the oxygen concentration in the samples under investigation corresponds to the stoichiometry $\rm YBa_2CuO_3O_{6.93\pm 0.02}$.

The density of samples determined by hydrostatic weighing was found to be 5.70 $g/cm^3$, i.e.,
90\% of the value obtained by x-ray investigations.

\begin{figure}[]
\includegraphics[width=8.5cm,angle=0]{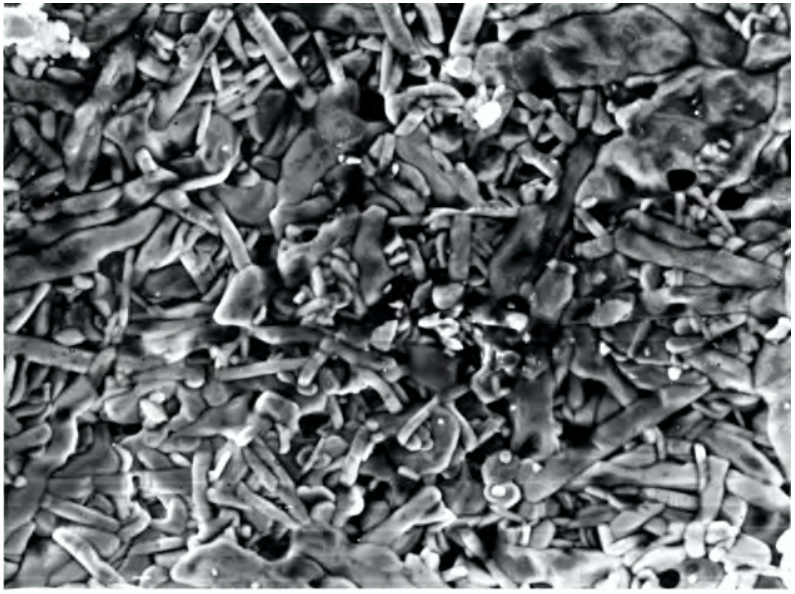}
\caption[]{Microstructure of the fracture surface of a  $\rm YBa_2Cu_3O_{7-\delta}$  pellet, observed under an electron microscope ($\times$2000).}
\label{Fig1}
\end{figure}

The microstructure of the samples, determined by the electron microscopy technique on JSM-120 (Japan) set-up, is presented in Fig. \ref{Fig1}. The characteristic size of the crystallites was found to be $\sim 5\times 5\times20~\mu m^3$.

\begin{figure}[]
\includegraphics[width=8cm,angle=0]{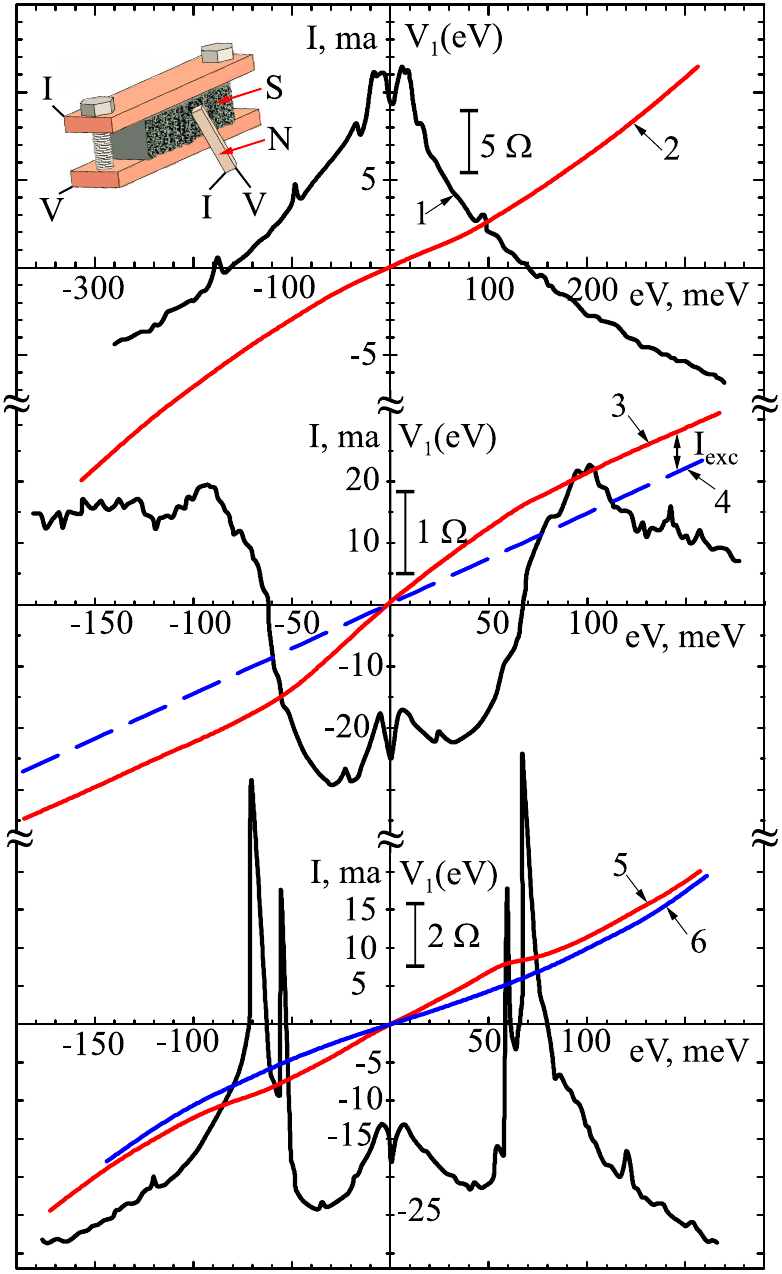}
\caption[]{Current-voltage characteristics and their first derivatives $V_1(eV)\sim dV/dI$ for some point contacts $\rm YBa_2Cu_3O_{7-\delta}-Ag$ ($Cu$ for curves 1 and 2):
1,2) $R_0(V=0)=30~\Omega, T=4.2~K; \  3,4) R_N\simeq 6.3~\Omega, T=4.2~K; \  5,7) R_0(V=0)=6.9~\Omega, T=4.2~K; \  6) T=82~K$.}
\label{Fig2}
\end{figure}

We studied the PC characteristics ($I(eV)$ and $V_1(eV)\sim dV/dI$) of junctions of the type superconductor -
normal metal, in which high-purity copper or silver was used as the normal metal. By measuring the $V_1(eV)$ and $I(eV)$ dependences over a wide range of temperature, we were able to determine the local value of the critical temperature $T_c$ at the point of contact of the electrodes. These values were found to be in the interval 78-87~$K$, i.e., slightly lower than the value $T_c(M)\simeq 93~K$ for the bulk samples (see inset to Fig. \ref{Fig2}).

Following the works of Artemenko, Volkov, and Zaitsev \cite{4}, and of Tinkham et al.\cite{5}, we have a clear understanding of the nature of excess current in superconducting PC satisfying the condition
$d<\xi(T)\ (1-T/T_c)^{1/4}$ ($d$ is the construction size and $\xi(T)$ is the coherence length), and also of the effect of elastic scatterers on the excess current in the constriction region \cite{6}. It was found that the diagnostic capabilities of point contacts in determining the energy gap in a superconductor are no worse than those of tunnel junctions. As far as the low-dispersion materials like the new superconductors are concerned, point contacts are even more advantageous since the probing region for them is smaller than for conventional tunnel junctions.

The original samples of yttrium ceramics in the form of cold-pressed pellets ($\O=10~mm, t=1.5~mm$) and bars of normal metal ($1\times 1\times 10~mm^3$) were fastened in a special measuring unit, which was then immersed in a helium cryostat. A remote drive mechanism was used to fracture the pellets directly in liquid helium, and a second electrode was brought close to the fracture surface until a weak electric contact was obtained (see inset to Fig. \ref{Fig2}). The contact with electrodes was established with the help of soft spring dampers to prevent a noticeable deformation or damage to the electrodes. Moreover, the measuring unit allowed for the displacement of the point of contact of the probe electrode with the fracture surface of the ceramic pellet. A brief exposure (for a few hours) of the pellets removed from the cryostat to air did not noticeably deteriorate the superconducting properties at the fracture surface if the temperature of the extracted samples was $\gtrsim 300~K$.

Various types of PC characteristics could be recorded at different points of contact of ($\rm Cu$ or $\rm Ag$) probe electrodes with the fracture surface of the ceramic pellet. One of the most frequently encountered characteristics is presented in Fig. \ref{Fig2} (curves 1 and 2). It is characterized by the presence of a structureless minimum on $V_1(eV)\sim dV/dI$ (curve 1) at $V=0$ (caused by the onset of superconductivity) and a semiconductor-type conductivity at large bias voltages. As far as the sharp peaks in the differential resistance $R_D$ on $V_1(eV)$ (which are encountered so frequently in the investigated point contacts) are concerned, their number, shape and position on the energy axis vary over a wide range for different contacts. With increasing temperature, these peaks are displaced towards lower energies, their intensity gradually decreases, and they disappear either together with the central minimum (near $V=0$) at $T=T_c$, or at a lower temperature. There was practically no excess current $I_{exc}$ in these contacts, which reflects a very low concentration of the superconducting phase in the constriction region. Typical resistance values of these contacts varied from a few tens to a few hundreds ohms. The contacts corresponding to the upper limit of this interval often revealed the absence of the central minimum and the peaks on $V_1(eV)$ (for point contacts with higher resistance values, this situation was always observed). This points towards the existence of a thin nonsuperconducting layer on the fracture surface of the yttrium ceramic.

Point-contact characteristics with a relatively high $I_{exc}$ (Fig. \ref{Fig2} curve 3) and clearly manifested minima on $V_1(eV)$ (curve 4) in the region of growing $I_{exc}$ correspond to contacts with a lower resistance than in the case considered above (the resistance varied from a few ohms to a few tens of ohms). According to the theory \cite{6} for N-c-S contacts (c stands for constriction), the position of such peaks on the energy axis corresponds to the values of the energy gap $\Delta$ at the S-electrode, if the electron parameters of both faces are considerably different from one another or if the constriction region contains elastic scatterers of electrons.

According to curves 3 and 4 in Fig. \ref{Fig2}, the conductivity of the contacts for $eV>\Delta$ is of metal type. However, there is an equal probability of observing the semiconductor type of resistivity in other contacts (curves 5 and 7). For both types of contacts, minima of different intensities are observed on $V_1(eV)$ dependence near $V = 0$. These minima could be suppressed by a magnetic field or microwave radiation at liquid helium temperature. This suppression did not affect the gap minima on $V_1(eV)$ in any way. Apparently, the nature of the central minimum in this type of contacts is associated with the proximity effect, i.e., with the induction of superconductivity in the normal probe electrode. This is confirmed by the disappearance of the central minimum upon heating the sample to a temperature $T\ll T_c$.

Sometimes, the suppression of the central minimum on $V_1(eV)$ near $eV = 10-15~meV$ is followed by the emergence of an additional minimum (or a point of inflection), which can be associated with the existence of a second (smaller) gap in yttrium ceramics, as in the case of lanthanum ceramics \cite{2}.

Moreover, some contacts showed $R_D$ peaks of different intensities (similar in nature and analogous to the $R_D$ peaks observed for the above-mentioned contacts). This can be attributed to a nonuniform distribution of the critical parameters of a superconductor over the volume of the PC constriction \cite{7}. Point contacts with a relatively large value of $I_{exc}$ exhibit a clear correlation between the positions of the $R_D$ peaks on the energy axis and the stepwise decrease in $I_{exc}$, the extent of blurring of a step determining the shape and intensity of the jump in $R_D$. (For example, the blurred peak in the region of $130-140~meV$ on curve 4 in Fig. \ref{Fig2} corresponds to an insignificant smooth decrease in the value $I_{exc}$).

\begin{figure}[]
\includegraphics[width=8cm,angle=0]{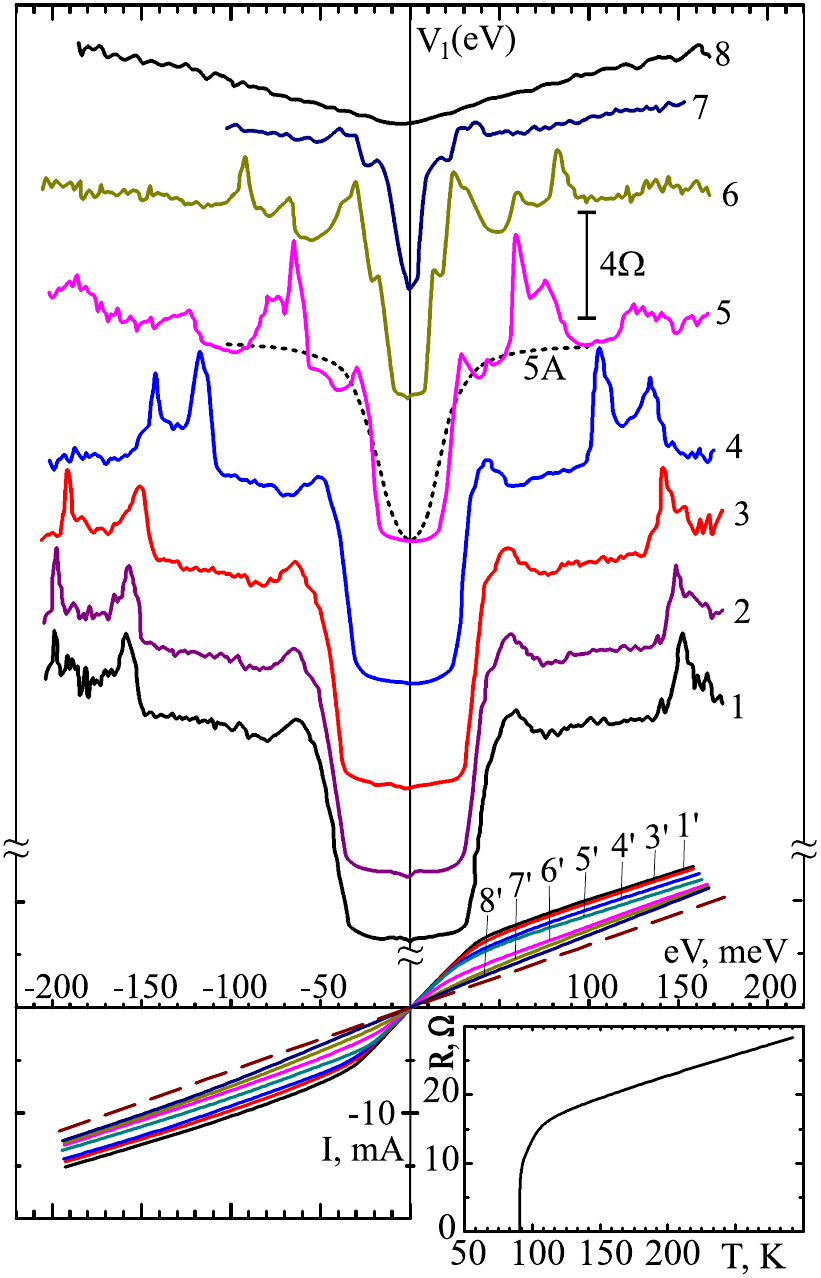}
\caption[]{Effect of temperature on the dependences ($V_1(eV$) (curves
1-8) and $I(eV)$ (curves 1'-8'); $ R_N\simeq 9~\Omega: 1,1') T=4.2~K;\  2) T=10~K;\
3,3') T=20.3~K; \ 4,4') T=49.9~K;\   5,5') T=69.7~K;\  6,6') T=75.6~K;\
7,7') T=78~K; \  8,8') 80.9~K$. The dotted curve 5A is the dependence $dV/dI(eV)$ calculated for $T=69.7~K,\  \Delta=11~meV$. The dashed line is parallel to the IVC (curve 1') in the energy interval 50-100~$meV$. The inset shows the $R(T)$ dependence for a bulk sample.}
\label{Fig3}
\end{figure}

The highest values of $I_{exc}$ (in our work) were observed in point-contacts having a resistance less than $20~\Omega$ for the case when $V_1(eV)$ curves did not contain any energy gap minima (Fig. \ref{Fig3}). It is well known \cite{6} that such characteristics must be formed not only when the point-contact constriction does not contain a barrier layer reflecting the electrons, but also when the electron parameters $p_F$ and $v_F$ are equal at the faces. The possibility of realizing such a situation in the investigated contacts can be associated with the existence of preferred crystallographic directions in the yttrium ceramic with a high electric conductivity, which may coincide with the point-contact axis (i.e., with the direction of current flow) and give rise to the observed characteristics.

The fact that the central minimum (near $V =0$) on $V_1(eV)$ (Fig. \ref{Fig3}, curve 1) has a very low intensity for this (third) type of contacts is worth noting. In the same way as for the previous type of contacts, this minimum is more strongly affected by temperature, magnetic field and microwave radiation than the main gap peculiarities. If the minimum under consideration is indeed caused by the induced superconductivity in the normal probe electrode, its low intensity may be due to an increase in the ratio $\xi(T)/d$ for highly conducting crystallographic directions.

An analysis of the nature of variation of the temperature dependences of $V_1(eV)$ for the last type of contacts (Fig. \ref{Fig3}) shows that the temperature blurring of the gap singularity for $T\gg 4.2~K$ (especially in the vicinity of $T_c$) is much smaller than the theoretical value for N-c-S contacts \cite{6}. (Curves 5 and 5A in Fig. \ref{Fig3} show the experimental and theoretical dependences $V_1(eV)$ for $T\simeq 70~K$.)

This discrepancy may be associated with the existence of a narrow peak in the electron density of states near the Fermi level in a yttrium ceramic.
It should be noted that on account of the existence of granules on the surface of the conducting layer with a suppressed order parameter (let us denote it by N'), the formation of PC characteristics will depend significantly on the Andreev reflection processes at the N'-S boundary if the contact diameter \emph{d} exceeds the thickness of the N'-layer. (In this case, the N-N' boundary will not significantly affect the PC characteristics.) Assuming further that the narrow peak in the density of states is preserved in the N'-layer as well, we can explain the extremely small blurring of the experimented dependences $V_1(eV)$ by temperature variations.

The possible displacement of the gap singularities on the $V_1(eV)$ dependence towards higher bias voltages by an amount equal to the voltage drop across the N'-layer for the low resistance
 ($R_N < 20~\Omega$) contacts at low temperatures can be obviously neglected since the position of the gap singularities on the $eV$ axis did not depend on $R_N$ for these contacts.

 However, the situation may change radically at higher temperatures. As a matter of fact, the effectiveness of heat removal is considerably reduced at $T\gg 4.2~K$ under conditions of a significant energy release (especially in low-resistance contacts). This may result in a local heating of the superconducting face (above $T_c)$ and in a corresponding displacement of the N'-S boundary away from the center of the contact. Since $I_{exc}$ is maintained up to temperatures $T \gtrsim 80~K$, this shift cannot be significant in comparison with the contact diameter \emph{d}.

 Nevertheless, the formation of the $V_1(eV)$ dependence at high temperatures and, in particular, the formation of a "flat bottom" at $eV \leq \Delta$, as well as the noncoincidence of the corresponding theoretical and experimental fragments of $V_1(eV)$, may depend to a large extent on the above-mentioned violation of superconductivity at small distances from the N'-S boundary. (This seems to be possible in view of small coherence lengths in the investigated superconductor.)

It was mentioned above that the absence of a minimum in the $V_1(eV)$ dependence at $eV =\Delta$ (Fig. \ref{Fig3}) may be attributed to a small resistivity $\rho_c$ along certain crystallographic directions in the investigated ceramic $\rm YBa_2Cu_3O_{7-\delta}$. Let us estimate roughly the contact
resistivity  $\rho_c$ by considering the example of the contact shown in Fig. \ref{Fig3}. Since the contact resistance in the vicinity of $V = 0$ is almost completely determined by the normal face which is assumed to be clean while the filling of the ceramic face by the superconducting phase is assumed to be complete, the contact diameter $d = (16\rho l/3\pi R_0)^{1/2}$. Taking $\rho l\simeq 0.9\cdot 10^{-11}~\Omega\cdot cm^2$    for \emph{Ag} and the contact resistance near $V =0$	 to be $R_0 \simeq 3~\Omega$, we obtain $d\simeq$220~\AA. (In the method used here, the thickness of the N'-layer with a suppressed order parameter must be assumed to be much smaller than $d$.) Further, we assume that the electron mean free paths \emph{l} in the ceramic are short, i.e., the diffusion regime of current flow is realized in the superconducting face. This gives the following value for $\rho_c:  \rho_c= (R_N - R_0)\cdot d\simeq 13\cdot 10^{-6}~\Omega\cdot cm$, where $R_N\simeq 9~\Omega$ is the contact resistance for $eV > \Delta$. (The corresponding critical current density is found to be $\sim 10^9~A/cm^2$.)

It should be noted that the value of $\rho_c$ obtained above is just the upper limit of this quantity, since $l$, may not be very small in this case. This, in turn, may lead to smaller values of $\rho_c$.

The existence of clearly manifested minima in the first derivatives $V_1(eV)\sim dV/dI$ of the IVC for contacts of the preceding type (curves 4 and 7 in Fig. \ref{Fig2}) points towards a lower conductivity of the ceramic face as compared to the normal face, or towards a contamination of the PC constriction region. It is not possible to calculate the diameter of such contacts since no data are available for this ceramic concerning the difference in electron parameters in different crystallographic directions, and also because the orientation of the contact axes relative to these crystallographic directions is not known.

For high-resistance point contacts (curves
1	and 2, Fig. \ref{Fig2}), the upper limit of d can be estimated from Maxwell's formula $d\sim\rho/R_N$ if we assign the highest known value $\rho_0\sim 10^{-3}~\Omega\cdot cm$ to the ceramic face. This gives $d\sim$1000~\AA\ for $R_N\sim 100~\Omega$. Since gap singularities and $I_{exc}$ on the PC characteristics are obviously not manifested for such contacts, it can be assume that the thickness of the degraded N'-layer in these cases is close to 1000~\AA.

The maximum value of the energy gap in the investigated ceramic $\rm YBa_2Cu_3O_{7-\delta}$ was found in our experiments to be $\Delta\simeq 40~meV$, which is in accord with the ratio $2\Delta/kT_c\simeq 12$. It was mentioned above that a second gap, which was much smaller than the main gap, was also observed for a number of contacts. The temperature dependence $\Delta(T)$ reveals a considerable departure from the BCS theory, as also the dependence for the earlier investigated lanthanum ceramic \cite{2} $\rm La_{1.8}Sr_{0.2}CuO_{4-\delta}$. (Investigations of this material are still being carried out.)

Finally, it should be emphasized that the values of $\Delta$ presented in this paper were measured for contacts with a high degree of filling of the PC region by the superconducting phase of homogeneous composition and are therefore quite reliable. It is also worthwhile to note that the results of this paper point towards the possibility of existence of certain crystallographic directions with a high electric conductivity in the ceramic. These directions are characterized by the appearance of a narrow peak in the density of states near the Fermi level.

The authors are deeply indebted to B.I. Verkin for encouragement and continued interest in this work.\vspace{\baselineskip}
\\
\textbf{NOTATION}\vspace{\baselineskip}
\\
Here $V_1(eV)$ is the effective voltage of modulating signal first harmonic, which is proportional to the first derivative $dV/dI$ of the current-voltage characteristic, $R_0$ the zero bias resistance of a point-contact in superconducting state, $R_N$ the zero-bias resistance of a point contact in the normal state, $R_D$ the differential resistance, $\Delta$ the superconducting energy gap, $d$ the point contact diameter, $\xi$ the coherence length, $I_{exc}$ the excess current, $p_F$ and $v_F$ the Fermi momentum and velocity respectively, $l$ the electron mean free path, and $\rho$ the resistivity of ceramics.

\end{document}